\documentclass[twocolumn,aps,prl,amsmath,amssymb]{revtex4-1}
\usepackage[latin9]{inputenc}
\usepackage{mathrsfs}
\usepackage{amsbsy}
\usepackage{amstext}
\usepackage{amssymb}
\usepackage{graphicx}
\usepackage{esint}

\makeatletter

\newcommand*\LyXThinSpace{\,\hspace{0pt}}

\usepackage{epsfig}
\usepackage{bm}
\usepackage{dcolumn}
\usepackage{color}

\makeatother

\begin{document}

\title{Effective theory for ultracold strongly interacting fermionic atoms
in two dimensions}

\author{Fan Wu$^{1}$}

\thanks{These authors contributed equally to this work.}

\author{Jianshen Hu$^{1}$}

\thanks{These authors contributed equally to this work.}

\author{Lianyi He$^{1,2}$ }

\author{Xia-Ji Liu$^{3}$ }

\author{Hui Hu$^{3}$}

\affiliation{$^{1}$Department of Physics and State Key Laboratory of Low-Dimensional
Quantum Physics, Tsinghua University, Beijing 100084, China}

\affiliation{$^{2}$Collaborative Innovation Center of Quantum Matter, Beijing
100084, China}

\affiliation{$^{3}$Centre for Quantum and Optical Science, Swinburne University
of Technology, Melbourne, Victoria 3122, Australia}

\date{\today}
\begin{abstract}
We propose a minimal theoretical model for the description of a two-dimensional
(2D) strongly interacting Fermi gas confined transversely in a tight
harmonic potential, and present accurate predictions for its zero-temperature
equation of state and breathing mode frequency based on existing auxiliary-field
quantum Monte Carlo data. We show that the minimal model Hamiltonian
needs at least two independent interaction parameters, the 2D scattering
length and effective range of interactions, in order to quantitatively
explain recent experimental measurements with ultracold 2D fermions.
We resolve in a satisfactory way the puzzling experimental observations
of the smaller than expected equations of state and breathing mode
frequency. Our establishment of the minimal model for 2D fermions
is crucial to understanding the Berezinskii-Kosterlitz-Thouless transition
in the strongly correlated regime.
\end{abstract}


\maketitle
Two-dimensional (2D) quantum many-body systems are of great interest,
due to the interplay of reduced dimensionality and strong correlation,
which leads to enhanced quantum and thermal fluctuations \cite{Randeria1989}
and a number of ensuing quantum phenomena such as Berezinskii\textendash Kosterlitz\textendash Thouless
(BKT) physics \cite{Berezinskii1972,Kosterlitz1973}. In this respect,
the recently realized 2D Fermi gas of ultracold $^{6}$Li and $^{40}$K
atoms under a tight axial confinement provides a unique platform \cite{Levinsen2015,Turlapov2017},
with unprecedented controllability particularly on interatomic interactions.
To date, many interesting properties of ultracold 2D Fermi gases have
been thoroughly experimentally explored \cite{Turlapov2017}, including
the equation of state (EoS) at both zero temperature \cite{Makhalov2014,Martiyanov2016}
and finite temperature \cite{Fenech2016,Boettcher2016}, radio-frequency
spectroscopy \cite{Frohlich2011,Sommer2012,Zhang2012}, pair momentum
distribution \cite{Ries2015}, first-order correlation function and
BKT transition \cite{Murthy2015}, and quantum anomaly in breathing
mode frequency \cite{Vogt2012,Holten2018,Peppler2018}. These results
may shed light on understanding other important strongly correlated
2D systems, such as high-$T_{c}$ layered cuprate materials \cite{Loktev2001},
$^{3}$He submonolayers \cite{Ruggeri2013}, exciton-polariton condensates
\cite{Deng2010} and neutron stars \cite{Pons2013}.

The present theoretical model of ultracold 2D Fermi gases is simple
\cite{Levinsen2015,Turlapov2017}. Under a tight harmonic confinement
with trapping frequency $\omega_{z}$ along the axial $z$-axis and
a weak confinement $\omega_{\perp}$ in the transverse direction,
the kinematic 2D regime is reached when the number of atoms $N$ is
smaller than a threshold $N_{2D}\simeq(\omega_{z}/\omega_{\perp})^{2}$,
so all the atoms are forced into the ground state of the motion along
$z$ \cite{Turlapov2017}. The interatomic interactions are then described
by a \emph{single} $s$-wave scattering length $a_{2D}$ \cite{Makhalov2014},
which is related to a 3D scattering length $a_{3D}$ via the quasi-2D
scattering amplitude \cite{Petrov2001}. Various experimental data
have been compared and benchmarked with different theoretical predictions
of the simple 2D model \cite{Bertaina2012,Orel2011,Hofmann2012,Taylor2012,Bauer2014,Barth2014,He2015,Shi2015,Mulkerin2015,Anderson2015}.
For EoS, i.e., the chemical potential and pressure at essentially
zero temperature, good agreements were found \cite{Makhalov2014,Boettcher2016}.
But, at the \emph{quantitative} level the experimental data somehow
lie systemically below the accurate predictions from auxiliary-field
quantum Monte Carlo (AFQMC) simulations \cite{Makhalov2014,Boettcher2016}.
The discrepancy is not so serious and might be viewed as an indicator
of small deviation from the 2D kinematics \cite{Turlapov2017}, in
spite of the fact that the 2D condition $N\ll N_{2D}$ is well satisfied.
However, a serious problem does arise when two experimental groups
measured the breathing mode frequency in the deep 2D regime most recently
\cite{Holten2018,Peppler2018}. The observed frequency turned out
to be much smaller than the well-established theoretical prediction
in the strongly interacting regime \cite{Hofmann2012,Taylor2012}.
This discrepancy is at the \emph{qualitative} level, suggesting that
the simple 2D model with a single parameter $a_{2D}$ may not be sufficient
for the description of ultracold 2D Fermi gases \cite{Hu2019}.

The purpose of this Letter is to provide a minimal theory of ultracold
2D Fermi gases, with the inclusion of a \emph{properly} defined effective
range of interactions (see Fig. 1). The significant role played by
effective range was realized in our previous work \cite{Hu2019}.
However, the determination of the effective range there turns out
be problematic. We solve the proposed model Hamiltonian at zero temperature
by taking into account strong pair fluctuations at Gaussian level
and beyond (Fig. 2), with the help of a correlation energy from AFQMC
in the zero-range limit \cite{Shi2015}. This enables us to predict
accurate EoS (Fig. 3 and Fig. 4), as well as reliable breathing mode
frequency (Fig. 5). The puzzling quantitative and qualitative discrepancies,
observed in the previous comparisons between experiment and theory
\cite{Turlapov2017,Makhalov2014,Boettcher2016,Holten2018,Peppler2018},
are therefore naturally resolved in a satisfactory way.

\textit{Effective range of interactions}. We start by considering
the collision of two fermions with mass $M$ and unlike spin in a
highly anisotropic harmonic trapping potential, described by a quasi-2D
scattering amplitude \cite{Petrov2001},
\begin{equation}
f_{Q2D}(k;a_{3D},a_{z})=\frac{4\pi}{\sqrt{2\pi}a_{z}/a_{3D}+\varpi\left(k^{2}a_{z}^{2}/2\right)},\label{eq:fQ2D}
\end{equation}
where $a_{z}\equiv\sqrt{\hbar/(M\omega_{z})}$ is the harmonic oscillator
length along the $z$-axis and the function $\varpi(x)$ has the expansion
$\varpi(x\rightarrow0)\simeq-\ln(2\pi x/\mathcal{B})+(2\ln2)x+i\pi$
with $\mathcal{B}\simeq0.905$ \cite{Petrov2001}. In the simplest
treatment, one may parameterize the quasi-2D collision using a 2D
scattering length $a_{2D}$ \cite{Turlapov2017,Makhalov2014}, by
setting the 2D scattering amplitude $f_{2D}(k;a_{2D})=-2\pi/\ln[ka_{2D}(k)/i]=f_{Q2D}(k;a_{3D},a_{z})$.
In general, one thus obtains a momentum-dependent $a_{2D}(k)$, which
in the zero-energy limit takes the form $a_{2D}(k\rightarrow0)=a_{s}\equiv a_{z}\sqrt{\pi/\mathscr{\mathcal{B}}}\exp(-\sqrt{\pi/2}a_{z}/a_{3D})$
\cite{Petrov2001}. The advantage of this simple treatment is that
the description \emph{universally} depends on a single parameter $a_{2D}$,
to be evaluated at a characteristic collision momentum $k_{0}$, i.e.,
$k_{0}=\sqrt{2M\tilde{\mu}}/\hbar$, where $\tilde{\mu}$ is the chemical
potential that does not include the two-body binding energy \cite{Turlapov2017,Makhalov2014}.

\begin{figure}
\centering{}\includegraphics[width=0.45\textwidth]{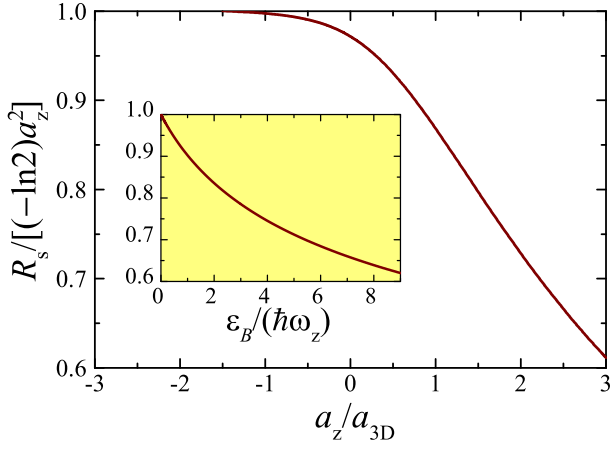}\caption{\label{fig1_Rseff} Confinement-induced effective range of interactions
$R_{s}$, in units of $R_{s}^{(0)}=(-\ln2)a_{z}^{2}$, as a function
of the inverse 3D scattering length $a_{z}/a_{3D}$. The inset shows
the effective range as a function of the two-body binding energy.}
\end{figure}

A more adequate parametrization of the 2D collision is to include
an effective range of interactions $R_{s}$ in the 2D scattering amplitude
\cite{Adhikari1986},

\begin{equation}
f_{2D}\left(k;a_{s},R_{s}\right)=\frac{4\pi}{-2\ln\left(ka_{s}\right)-R_{s}k^{2}+i\pi},\label{eq:fQ2DLowEnergy}
\end{equation}
whose pole gives a two-body bound state with binding energy $\varepsilon_{B}=\hbar^{2}\kappa^{2}/M$,
where the wavevector $\kappa$ satisfies $R_{s}=2\ln(\kappa a_{s})/\kappa^{2}$.
The same two-body bound state should be supported by the pole of the
quasi-2D scattering amplitude in Eq. (\ref{eq:fQ2D}) as well. By
setting $k\rightarrow i\kappa$ there, we find $\sqrt{2\pi}a_{z}/a_{3D}+\varpi[-\varepsilon_{B}/(2\hbar\omega_{z})]=0$
\cite{NoteEB}. Therefore, we can directly calculate the effective
range $R_{s}$, once $\varepsilon_{B}$ or $\kappa$ is solved at
a given $a_{z}/a_{3D}$. 

The effective range obtained in this way is reported in Fig. \ref{fig1_Rseff}.
It decreases monotonically from $R_{s}^{(0)}\equiv(-\ln2)a_{z}^{2}$
with increasing $a_{z}/a_{3D}$ (main figure) or binding energy $\varepsilon_{B}$
(inset). We note that $R_{s}^{(0)}$ can be easily derived from the
second expansion term in $\varpi(x\rightarrow0)$ and its magnitude,
i.e., $R_{s}^{(0)}\sim a_{z}^{2}$, is a clear indication of the quasi-2D
nature of atom collisions \cite{Turlapov2017,Petrov2001}. As the
wavefunction of two colliding atoms at distance within $a_{z}$ is
set by the full 3D contact interaction potential, these collisions
can never be \emph{purely} 2D. They can only be approximately treated
as 2D, out of the range $\sim a_{z}$.

\begin{figure}[t]
\begin{centering}
\includegraphics[width=0.45\textwidth]{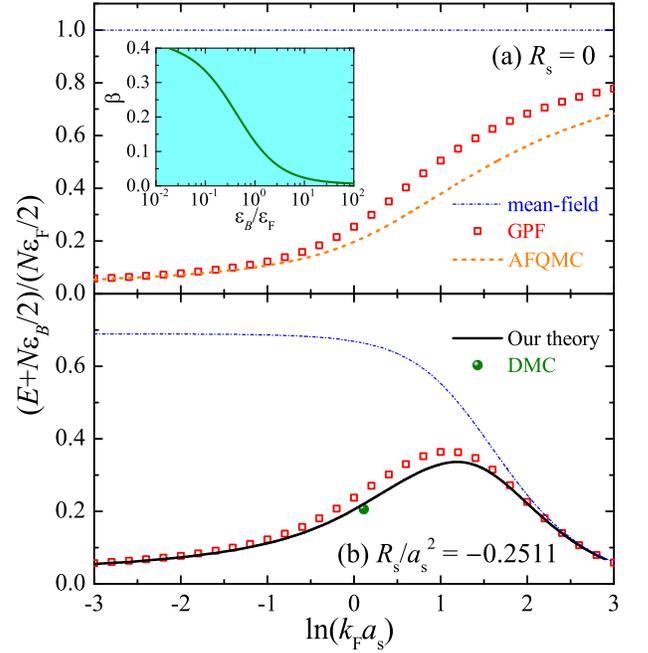} 
\par\end{centering}
\caption{\label{fig2_energy} Total energy with the two-body bound-state energy
subtracted as a function of $\ln(k_{F}a_{s})$, at $R_{s}=0$ (a)
and $R_{s}\simeq-0.2511a_{s}^{2}$ (b). The mean-field and GPF predictions
are shown by blue dot-dashed lines and red squares, respectively.
At zero range in (a), the latest AFQMC result \cite{Shi2015} is plotted
by orange dashed line. The inset shows the beta function $\beta=\Delta E_{\textrm{c}}/\Delta E_{\textrm{GPF}}$
(see Eq. (\ref{eq:energy})). At finite range in (b), our theory (black
solid line) is compared with the DMC data (green dot) \cite{Schonenberg2017}.}
\end{figure}

\textit{Many-body theory}. To account for the effective range $R_{s}$,
it is useful to adopt a two-channel model \cite{Hu2019,Liu2005,Schonenberg2017}:
\begin{eqnarray}
\mathcal{H} & = & \sum_{\mathbf{k}\sigma=\left\{ \uparrow,\downarrow\right\} }\xi_{\mathbf{k}}c_{\mathbf{k}\sigma}^{\dagger}c_{\mathbf{k}\sigma}+\sum_{\mathbf{q}}\left(2\xi_{\mathbf{q}/2}+\nu\right)b_{\mathbf{q}}^{\dagger}b_{\mathbf{q}}\nonumber \\
 &  & +\frac{g}{\sqrt{\mathcal{S}}}\sum_{kq}\left(b_{\mathbf{q}}c_{\mathbf{q}/2+\mathbf{k}\uparrow}^{\dagger}c_{\mathbf{q}/2-\mathbf{k}\downarrow}^{\dagger}+\textrm{h.c.}\right),
\end{eqnarray}
where $\xi_{\mathbf{p}}\equiv\hbar^{2}\mathbf{p}^{2}/(2M)-\mu$, and
$c_{\mathbf{k}\sigma}$ and $b_{\mathbf{q}}$ are the annihilation
operators of atoms and molecules in the open- and closed-channel,
respectively. The channel coupling $g$ is related to $R_{s}$, via
$R_{s}=-4\pi^{2}\hbar^{4}/(M^{2}g^{2})$, the detuning $\nu$ of molecules
is tuned to reproduce the binding energy $\varepsilon_{B}$, i.e.,
$\nu=-\varepsilon_{B}+(g^{2}/\mathcal{S})\sum_{\mathbf{k}}[\hbar^{2}\mathbf{k}^{2}/M+\varepsilon_{B}]^{-1}$
\cite{Hu2019,Schonenberg2017}, and $\mathcal{S}$ is the area.

We solve the model Hamiltonian at different orders of approximation
at zero temperature. Formally, the ground-state energy $E$ may be
decoupled as,
\begin{equation}
E\left[\ln\left(k_{F}a_{s}\right),k_{F}^{2}R_{s}\right]=E_{\textrm{MF}}+\Delta E_{\textrm{GPF}}+\Delta E_{\textrm{c}},\label{eq:energy}
\end{equation}
where $k_{F}=(2\pi n)^{1/2}$ is Fermi wavevector and $\varepsilon_{F}=\hbar^{2}k_{F}^{2}/(2M)$
is Fermi energy for a system with number density $n$. The mean-field
(MF) theory provides the leading term $E_{\textrm{MF}}$, while the
major correction arising from strong pair fluctuations at Gaussian
level can be obtained by using the Gaussian pair fluctuation (GPF)
theory \cite{He2015,Hu2019,Hu2006,Hu2007,Diener2008}, i.e., $\Delta E_{\textrm{GPF}}=E_{\textrm{GPF}}-E_{\textrm{MF}}$.
The effect of pair fluctuations \emph{beyond} Gaussian level may be
characterized by a correlation energy $\Delta E_{\textrm{c}}$, which
is anticipated to be much smaller than $\Delta E_{\textrm{GPF}}$.
To see this, in Fig. \ref{fig2_energy}(a) we plot the ground-state
energy in the zero-range limit ($R_{s}=0$), predicted by mean-field
theory, GPF theory \cite{He2015} and AFQMC simulation \cite{Shi2015}.
Indeed, the correlation energy given by the difference between the
GPF and AFQMC energies is notably smaller than $\Delta E_{\textrm{GPF}}$.
In particular, $\Delta E_{\textrm{c}}$ becomes vanishingly small
in the tight-binding limit of $\ln(k_{F}a_{s})\rightarrow-\infty$
\cite{He2015}. It is then useful to define a beta function $\beta=\Delta E_{\textrm{c}}/\Delta E_{\textrm{GPF}}\ll1$,
which varies as functions of the two dimensionless interaction parameters
$\ln(k_{F}a_{s})$ and $k_{F}^{2}R_{s}$. For small $k_{F}^{2}R_{s}$,
however, it seems plausible to assume that $\beta$ relies on $\varepsilon_{B}/\varepsilon_{F}$
only, whose dependence can be readily extracted in the zero-range
limit using the AFQMC data, as shown in the inset of Fig. \ref{fig2_energy}(a).
Other possible choice of the $\beta$-function is considered in Supplemental
Material \cite{SM}. 

We thus establish a viable procedure to calculate the ground-state
energy at nonzero effective range. For a given set ($k_{F}a_{s}$,
$k_{F}^{2}R_{s}$), we first calculate the binding energy $\varepsilon_{B}/\varepsilon_{F}$
and determine the value of $\beta$. Both mean-field and GPF theories
are then applied to obtain $E_{\textrm{MF}}$ and $\Delta E_{\textrm{GPF}}$,
and consequently $\Delta E_{\textrm{c}}=\beta\Delta E_{\textrm{GPF}}$.
In Fig. \ref{fig2_energy}(b), we present $E=E_{\textrm{GPF}}+\Delta E_{\textrm{c}}$
in black line for a fixed ratio $R_{s}/a_{s}^{2}\simeq-0.2511$, at
which we may benchmark our prediction against available high-precision
diffusion Monte Carlo (DMC) data (i.e., the single green dot) \cite{Schonenberg2017,NoteDMC}.
We find that the correction $\Delta E_{\textrm{GPF}}$ becomes smaller
at nonzero effective range. Towards the non-interacting limit ($a_{s}\rightarrow\infty$)
and hence large $k_{F}^{2}R_{s}$, $\Delta E_{\textrm{GPF}}$ vanishes
quickly. This is understandable, since pair fluctuations become weaker
with decreasing channel coupling $g$ and even mean-field theory may
provide accurate prediction at sufficiently large $k_{F}^{2}R_{s}$
\cite{Schonenberg2017}. The correlation energy also significantly
reduces at finite effective range and we find $\left|\Delta E_{\textrm{c}}\right|<0.02N\varepsilon_{F}$
at all interaction strengths for $R_{s}/a_{s}^{2}\simeq-0.2511$.
The agreement between our theory with DMC is excellent, with a difference
less than $0.01N\varepsilon_{F}$.

\begin{figure}[t]
\centering{}\includegraphics[width=0.48\textwidth]{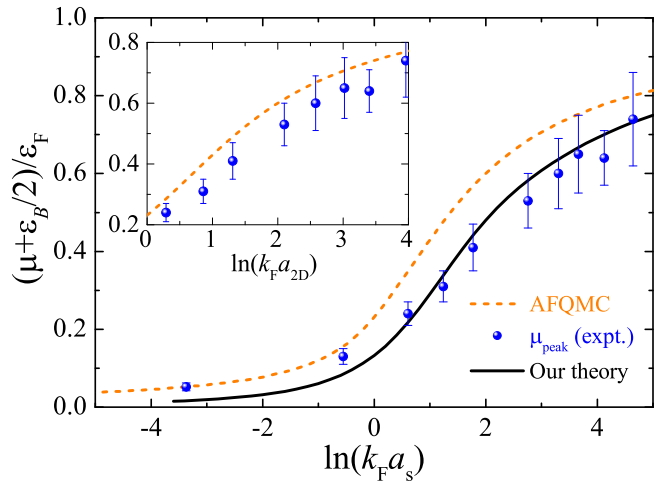}\caption{\label{fig3_mu} Chemical potential with the two-body bound state
contribution subtracted, as a function of $\ln(k_{F}a_{s})$ at the
number of atoms $N=N_{2D}$. The predictions of AFQMC (i.e., for zero
effective range) \cite{Shi2015} and our theory with a realistic effective
range as in the experiment \cite{Boettcher2016} are shown by orange
dashed line and black solid line, respectively, and are compared with
the experimental data for $\mu_{\textrm{peak}}$ (blue circles) measured
at $N\simeq N_{2D}$ \cite{Boettcher2016,NoteEoS}. The inset shows
the chemical potential as a function of $\ln(k_{F}a_{2D})$, where
$a_{2D}$ is the effective scattering length adopted in the experiment
\cite{Boettcher2016}.}
\end{figure}

\textit{Equation of state}. Once the ground-state energy $E$ of a
uniform 2D Fermi gas is determined, we calculate directly the chemical
potential $\mu$ and pressure $P$ using standard thermodynamic relations.
Experimentally, these \emph{homogeneous} EoS can be extracted from
a low-temperature trapped Fermi gas, by using the local density approximation
\cite{Butts1997}, which assigns a local chemical potential $\mu(r)=\mu_{\textrm{peak}}-V(r)$
to each position $r$ in the potential $V(r)=M\omega_{\perp}^{2}r^{2}/2$.
Both the peak chemical potential $\mu_{\textrm{peak}}$ and the\emph{
in situ} density distribution $n(r)$ can be experimentally measured
\cite{Makhalov2014,Fenech2016,Boettcher2016}, from which one deduces
the homogeneous density EoS $n(\mu)$. By further using the force
balance condition \cite{Makhalov2014}, $\nabla P(r)=-n(r)\nabla V(r)$,
the homogeneous pressure EoS $P(n)$ can also be determined.

In Fig. \ref{fig3_mu}, we show the experimental data for the peak
chemical potential $\mu_{\textrm{peak}}$, measured at different magnetic
fields (i.e., $a_{3D}$) and hence at different $\ln(k_{F}a_{s})$
\cite{Boettcher2016,NoteEoS}. Our predictions for the peak chemical
potential, calculated under the same experimental condition, are plotted
by the black solid line. We find a good agreement between theory and
experiment at $\ln(k_{F}a_{s})>0$. Due to the large effective range
of interactions in the experiment (i.e., $k_{F}^{2}R_{s}\lesssim-1.2$
at $N\simeq N_{2D}$ \cite{Boettcher2016}), the zero-range predictions
from AFQMC appear to strongly over-estimate the chemical potential.
The use of an \emph{effective} $a_{2D}$ can not fully explain the
discrepancy (see the inset and also Fig. 1 in Ref. \cite{Boettcher2016}),
as we mentioned earlier. The discrepancy can also be hardly understood
by possible systematic effects such as finite temperature and the
failure of a 2D model due to a finite filling factor $N/N_{2D}$ \cite{SM,He2019}.

\begin{figure}[t]
\begin{centering}
\includegraphics[width=0.48\textwidth]{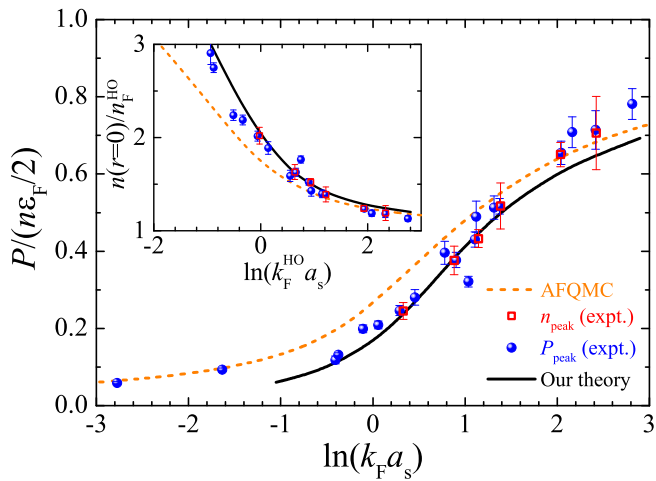} 
\par\end{centering}
\centering{}\caption{\label{fig4_pressure} Pressure as a function of $\ln(k_{F}a_{s})$
at $N=0.35N_{2D}$. We use blue circles and red squares to show the
experimental data from Ref. \cite{Makhalov2014} and Ref. \cite{Martiyanov2016}
with $N\simeq0.35N_{2D}$, respectively. The predictions of AFQMC
\cite{Shi2015} at zero range and our theory at finite range are shown
by orange dashed line and black solid line, respectively. Towards
the weakly interacting limit, the finite-temperature effect may become
sizable and up-shift the pressure data \cite{Makhalov2014}. The inset
shows the peak density (in units of $n_{F}^{\textrm{HO}}$) as a function
of $\ln(k_{F}^{\textrm{HO}}a_{s})$, where $n_{F}^{\textrm{HO}}$
and $k_{F}^{\textrm{HO}}=(2\pi n_{F}^{\textrm{HO}})^{1/2}$ are the
peak density and wave-vector of an ideal Fermi gas in traps.}
\end{figure}

In Fig. \ref{fig4_pressure}, we present the comparison between our
predictions and the experimental data \cite{Makhalov2014,Martiyanov2016}
for pressure at the trap center. In this case, we have $N\simeq0.35N_{2D}$
and therefore the effect of the effective range may become weaker.
Nevertheless, we can see clearly that in the strongly interacting
regime (i.e., $0<\ln(k_{F}a_{s})<2$), the experimental data lie systematically
below the zero-range results from AFQMC. The model Hamiltonian with
a finite effective range should be used, in order to quantitatively
understand the experimental measurement. We note that, in harmonic
traps the pressure at the center is fixed by the force balance condition
to $P=M\omega_{\perp}^{2}N/(2\pi$) \cite{Martiyanov2016}. Using
the peak density of an ideal trapped Fermi gas $n_{F}^{\textrm{HO}}=M\omega_{\perp}\sqrt{N}/(\pi\hbar)$
\cite{Turlapov2017}, we find that the peak density $n\equiv n(r=0)$
can be written in terms of the pressure at the trap center, i.e.,
$n/n_{F}^{\textrm{HO}}=[P/(n\varepsilon_{F}/2)]^{-1/2}$. This provides
an alternative way to illustrate the data, as shown in the inset.

In both Fig. \ref{fig3_mu} and Fig. \ref{fig4_pressure}, the agreement
between theory and experiment becomes worse at small $k_{F}a_{s}$,
suggesting the inadequacy of our theory towards the limit of a Bose-Einstein
condensate (BEC). This is because, experimentally the BEC regime is
reached by changing $a_{3D}$ instead of $k_{F}$. For a small positive
$a_{3D}$ the system is better viewed as a quasi-2D weakly interacting
BEC, with a 2D scattering length $a_{2D}^{(m)}$ determined from the
3D \emph{molecular} scattering length $a_{3D}^{(m)}\simeq0.6a_{3D}$
\cite{Petrov2004} and with an effective range $R_{s}^{(m)}\sim-a_{z}^{2}$.
Our two-channel model cannot fully recover this interaction-driven
BEC limit. For more details, we refer to Supplemental Material \cite{SM}.

\begin{figure}[t]
\begin{centering}
\includegraphics[width=0.48\textwidth]{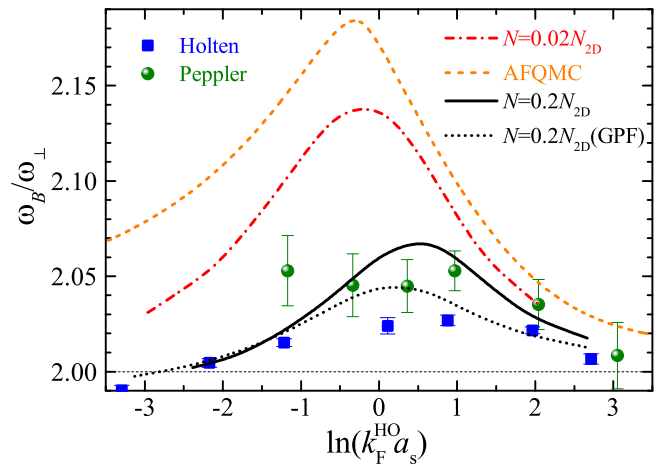} 
\par\end{centering}
\centering{}\caption{\label{fig4_QuantumAnomaly} Breathing mode frequency of 2D strongly
interacting fermions as a function of the interaction parameter $\ln(k_{F}^{\textrm{HO}}a_{s})$,
at different total number of atoms $N/N_{2D}\rightarrow0$ (AFQMC
\cite{NoteBMAFQMC}, orange dashed line), $0.02$ (red dot-dashed
line), and $0.2$ (black solid line) as in two recent experiments
by Holten \textit{et al.} \cite{Holten2018} at $0.10-0.18T_{F}$
(green circles) and Peppler \textit{et al.} \cite{Peppler2018} at
$0.14-0.22T_{F}$ (blue squares). Here $T_{F}$ is the Fermi temperature.
We show also the GPF prediction in the dotted line.}
\end{figure}

\textit{Breathing mode and quantum anomaly}. We now turn to consider
the breathing mode frequency, which was recently measured in two experiments
at $N\simeq0.2N_{2D}$ \cite{Holten2018,Peppler2018}, as shown in
Fig. \ref{fig4_QuantumAnomaly} by green circles and blue squares.
Theoretically, the zero-temperature breathing mode frequency can be
conveniently calculated by using the sum-rule approach \cite{Menotti2002,Hu2014},
\begin{equation}
\hbar^{2}\omega_{B}^{2}=-2\left\langle r^{2}\right\rangle \left[\frac{d\left\langle r^{2}\right\rangle }{d\left(\omega_{\perp}^{2}\right)}\right]^{-1},\label{eq:sumrule}
\end{equation}
where $\left\langle r^{2}\right\rangle =N^{-1}\int d^{2}\boldsymbol{r}[r^{2}n(r)]$
is the squared radius of the Fermi cloud at a given trapping frequency
$\omega_{\perp}$. In the classical treatment, a 2D Fermi gas is scale-invariant
\cite{Pitaevskii1997} and acquires a polytropic density EoS, $\mu(n)\propto n{}^{2}$.
As a result, the mode frequency is pinned to $2\omega_{\perp}$, regardless
of temperature and interactions \cite{Pitaevskii1997}. The deviation
of the breathing mode frequency away from $2\omega_{\perp}$ can be
viewed a quantum anomaly \cite{Hofmann2012,Taylor2012}, arising from
strong quantum pair fluctuations in 2D \cite{Olshanii2010}.

As readily seen from Fig. \ref{fig4_QuantumAnomaly}, the observed
quantum anomaly in the two experiments is far below the prediction
from AFQMC for zero-range interactions with a single 2D scattering
length \cite{NoteBMAFQMC}. It can only be understood when we use
the proposed minimal model for 2D ultracold fermions and take into
account the realistic finite effective range at $N\simeq0.2N_{2D}$.
The quantitative difference between our theory and experiment at $0<\ln(k_{F}a_{s})<1$
could be caused by the finite temperature in the two experiments \cite{SM},
which is in the range $[0.10-0.22]T_{F}$.

It turns out that the breathing mode frequency or quantum anomaly
depends sensitively on the effective range. The zero-range result
of AFQMC can hardly be asymptotically approached, even we decrease
the number of atoms down to just a few percent of $N_{2D}$ (see the
red dot-dashed line at $N=0.02N_{2D}$). In this case, however, the
deviation from the classical limit of $2\omega_{\perp}$ is very significant
and its experimental confirmation will deepen our understanding of
the long-sought 2D quantum anomaly in cold atoms \cite{Olshanii2010},
which was recently observed \cite{Murthy2019}.

\textit{Conclusions}. We have established a minimal model to describe
ultracold interacting fermions confined in two dimensions and have
solved it accurately at zero temperature with the help of existing
AFQMC results. We have shown that the confinement-induced effective
range of interactions has to be included, in order to understand the
recent measurements on quantum anomaly in a qualitative manner and
on equation of state at quantitative level. Our results pave the way
to investigate the crucial role played by effective range in other
two-dimensional quantum many-body systems and provide an excellent
starting point to address the fermionic Berezinskii-Kosterlitz-Thouless
transition with cold-atoms \cite{Murthy2015,Mulkerin2017}.
\begin{acknowledgments}
We thank Igor Boettcher for useful discussions. This research was
supported by the National Natural Science Foundation of China, Grant
No. 11775123 (L.H.), National Key Research and Development Program
of China, Grant No. 2018YFA0306503 (L.H.), and Australian Research
Council's (ARC) Discovery Program, Grant No. FT140100003 (X.-J.L),
Grant No. DP180102018 (X.-J.L), and Grant No. DP170104008 (H.H.).
\end{acknowledgments}


\begin{thebibliography}{10}
\bibitem{Note2DScatteringLength}The definition of a 2D scattering
length (denoted by $a_{2}$) used by Turlapov group in Ref. \cite{Makhalov2014}
and Ref. \cite{Martiyanov2016} is slightly different, and is related
to $a_{2D}(k_{0})$ by, $a_{2D}(k_{0})=(e^{\gamma_{E}}/2)a_{2}$,
where $\gamma_{E}\simeq0.577$ is Euler's constant and $k_{0}=\sqrt{2M\tilde{\mu}}/\hbar$.
We also note that, in Ref. \cite{Hu2019} we have used the notation
$a_{2D}=a_{2D}(k\rightarrow0)$, which is exactly $a_{s}$ defined
in this work.
\bibitem{NoteEB}This equation can be rewritten as, $a_{z}/a_{3D}=\mathcal{F}[\varepsilon_{B}/(\hbar\omega_{z})]$,
where the function $\mathcal{F}(x)$ is given by, $\mathcal{F}(x)=\int_{0}^{\infty}du(4\pi u^{3})^{-1/2}[1-e^{-xu}/\sqrt{(1-e^{-2u})/(2u)}]$.
\end{thebibliography}
\end{document}